\documentclass{aa}
\usepackage{amsmath}
\usepackage{graphicx}
\usepackage{txfonts}
\begin{document}

   \title{Impactor Flux and Cratering on the Pluto-Charon System}

   \author{ G. C. de El\'\i a  
           \thanks{gdeelia@fcaglp.unlp.edu.ar},
           R. P. Di Sisto
%          \thanks{}
          \and
           A. Brunini 
%	  \thanks{}
          }

   \offprints{G. C. de El\'{\i}a
    }

  \institute{Facultad de Ciencias Astron\'omicas y Geof\'\i sicas, Universidad 
Nacional de La Plata and \\
Instituto de Astrof\'{\i}sica de La Plata, CCT La Plata-CONICET-UNLP \\
   Paseo del Bosque S/N (1900), La Plata, Argentina.
                }

   \date{Received / Accepted}

%-------------------------------------------------------

\abstract
{}
{We study the impactor flux and cratering on Pluto and Charon due to the collisional evolution of Plutinos.
Plutinos are those trans-Neptunian objects located at $\sim$ 39.5 AU, in the 3:2 mean motion resonance with Neptune.
}
{To do this, we develop a statistical code that includes catastrophic collisions and cratering events, and takes
into account the stability and instability zones of the 3:2 mean motion resonance with Neptune.
Moreover, our numerical algorithm proposes different initial populations that account for the uncertainty in the size distribution 
of Plutinos at small sizes.
}
{Depending on the initial population, our results indicate the following. The number of $D >$ 1 km Plutinos streaking Pluto over 3.5
Gyr is between 1\,271 and 5\,552. For Charon, the number of $D >$ 1 km Plutino impactors is between 354 and 1\,545.
The number of $D > 1$ km craters on Pluto produced by Plutinos during the last 3.5 Gyr is between 43\,076 and 113\,879. For 
Charon, the number of $D > 1$ km craters is between 20\,351 and 50\,688.
On the other hand, the largest Plutino impactor onto Pluto has a diameter between $\sim$ 17 and 23 km, which produces a crater
with a diameter of $\sim$ 31 -- 39 km. In the same way, the largest Plutino impactor onto Charon has a diameter between $\sim$ 10 and 15 km, which
produces a crater with a diameter of $\sim$ 24 -- 33 km. Finally, we test 
the dependence of results on the number of Pluto-sized objects in the 
Plutino population. If 2 Pluto-sized objects are assumed in the 3:2 Neptune resonance, the total number of Plutino impactors both onto Pluto as 
Charon with diameters $D >$ 1 km is a factor of $\sim$ 1.6 -- 1.8 larger than that obtained considering only 1 Pluto-sized object in this resonant 
region. 
}
{Given the structure of the trans-Neptunian region, with its dynamically different populations, it is necessary 
to study in detail the contribution of all the potential sources of impactors on the Pluto-Charon system, to obtain
the main contributor and the whole production of craters. Then, we will be able to contrast those studies with 
observations which will help us to understand the geological processes and history of the surface of those worlds. 
}


\authorrunning{G. C. de El\'\i a, R. P. Di Sisto \&
               A. Brunini
               }
\titlerunning{
Impactor Flux and Cratering on the Pluto-Charon System
                             }

\maketitle
\section{Introduction}

Pluto and Charon are members of a vast population of icy bodies beyond Neptune and 
constitute in fact the first discovered binary trans-Neptunian object (TNO).
In the trans-Neptunian region there are at first four dynamical classes 
(Chiang et al. \cite{Chiang2007}). The Classical Objects those with semimajor axis $a$ greater 
than $\sim$42 AU and low eccentricity orbits, the Scattered Disk Objects (SDOs) 
with perihelion distance $q > 30$ AU with large eccentricities, the Resonant Objects those in mean 
motion resonance with Neptune and the Centaurs with perihelion distance $q < 30$ AU. 

The 3:2 mean motion resonance with Neptune, located at $\sim$ 39.5 AU, is the most 
densely populated one in the trans-Neptunian region. The residents of this resonant region are 
usually called Plutinos because of the analogy of their orbits with that of Pluto, 
which is its most representative member. Aside from Pluto and
its largest moon Charon, the Minor Planet Center (MPC) database contains $\sim$ 200 
Plutino candidates.

Pluto and Charon have been exposed to impacts with minor bodies as 
all the objects in our Solar System.  Cratering is one of the most important processes that determine the
morphology of the surface of a Solar System object. The understanding and quantification of the impactor source
population onto an object and the observation of the object surface help to understand the dynamical and 
physical history of both the impactor population and the target.

All of the detailed knowledge of the surface composition
of Pluto and Charon has been obtained from telescopic observations of the spectrum of
sunlight reflected from their surfaces. Pluto's reflectance spectrum shows 
absorption bands of methane ice, and an absorption band that could be related 
to the presence of CO and nitriles. The surface of Charon can be modeled by 
pure water ice darkened by a spectrally neutral continuum absorber 
(Protopapa et al. \cite{Protopapa2008}). 

 However, it will not be until 2015 that we will have a real idea of the morphology of the 
surfaces of Pluto and Charon, with the fly by of the Nasa's New Horizon Pluto-Kuiper Belt mission to Pluto system.
The New Horizons mission is the first one to the Pluto system and the Kuiper Belt 
and was launched on 19 January 
2006 on a Jupiter Gravity Assist trajectory toward the Pluto system for a 
14 July 2015 closest approach. It will study the Pluto system over a 5-month period beginning in early 2015 
in particular providing measurement of cratering records. 

The comparison of the predicted theoretical crater production from a 
given source with the observed 
surface of Pluto and Charon may account for the geological processes acting on the 
surface of the objects and if cratering collisions onto Pluto and Charon are an important 
surface modification process.  

Then, it is very important to study all the possible sources of crater production 
on Pluto and Charon in order to know the total crater production and to 
contrast them with observations.

The possible main contributors to the 
impactor flux on Pluto and Charon would be on one hand all type of comets.
This topic has been addressed by Weissman \& Stern (\cite{Weissman1994}), 
Durda \& Stern (\cite{Durda2000}), and Zahnle et al. (\cite{Zahnle2003}). 
In general comets have eccentric orbits that can cross Pluto's orbit when they 
enter to the planetary region from their source, either the trans-Neptunian zone or 
the Oort Cloud, or when they leave the inner Solar System because of the perturbations 
of the planets.

Weissman \& Stern (\cite{Weissman1994}) estimated current impact rates of comets
on Pluto and Charon. They showed that cratering on both bodies is dominated by Kuiper
Belt and inner Oort cloud comets.
Then, Durda \& Stern (\cite{Durda2000}) calculated collision rates in the Kuiper 
Belt and Centaur region through a numerical model. 
They estimated that the flux of Kuiper Belt projectiles 
onto Pluto and Charon is $\sim$ 3 -- 5 times that of Weissman \& Stern (\cite{Weissman1994}).
Later, Zahnle et al. (\cite{Zahnle2003}) studied the cratering rates for the moons of the jovian planets and
Pluto produced mainly by ecliptic comets, obtaining results consistent with the previous estimates.
 We will analyze those studies in a discussion section.

On the other hand, Plutinos may be the other important source of impactors on Pluto 
and Charon. Plutinos shared the same dynamical conditions that Pluto and Charon, in fact 
they are all located in the same 3:2 mean motion resonance with Neptune so they have 
a certain collision probability. Recently, de El\'{\i}a et al. (\cite{deElia2008}) 
performed a collisional evolution of Plutinos and determined collisional rates among 
these objects. That work allows us to study the impactor flux on Pluto and Charon 
due to the collisional evolution of Plutinos. 
Then, in this paper we are going to evaluate the contribution of Plutinos to the 
impactor flux and cratering on Pluto and Charon and also determine if the Plutino population can 
be considered a primary source of impactors on the Pluto-Charon system.

\section{The Full Model}

To simulate the collisional and dynamical evolution of the Plutino population, we use the statistical code 
developed by de El\'\i a et al. (\cite{deElia2008}). This algorithm considers catastrophic collisions and cratering events,
and takes into account the main dynamical characteristic associated to the 3:2 mean motion resonance with Neptune.
In the following, we give a brief description of the initial populations, the collisional parameters and the main 
dynamical considerations used in our model.

\begin{figure}
\centering
\resizebox{\hsize}{!}{\includegraphics{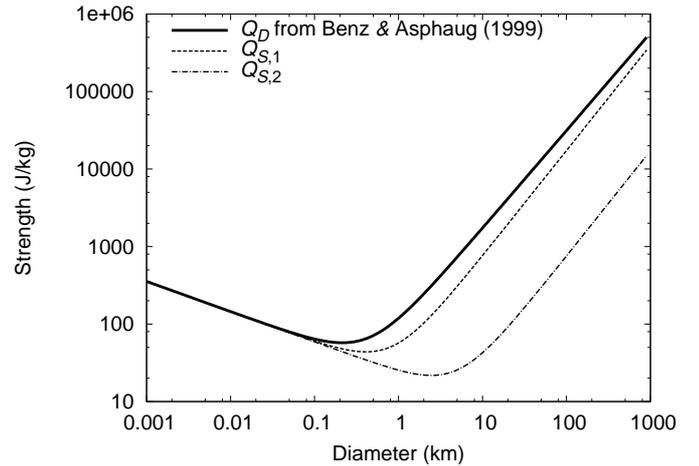}}
\caption{
Impact strength. The dashed lines represent the two different $Q_{S}$ laws used in our simulations. The $Q_{D}$ law from
Benz \& Asphaug (\cite{Benz1999}) for icy bodies at 3 km\,\,s$^{-1}$ is plotted as a solid line.
        }
\label{Fig1}
\end{figure}

\subsection{Initial Populations}

From Kenyon et al. (\cite{Kenyon2008}), the cumulative size distribution of the resonant population of the trans-Neptunian region shows a
break at a diameter $D$ near 40 -- 80 km. Moreover, for larger resonant objects, the population seems to have a shallow
size distribution with a cumulative power-law index of $\sim$ 3. 
From this, the general form of the cumulative initial population used in our model to study the
collisional and dynamical evolution of the Plutinos can be written as follows
\begin{eqnarray}
N(>D) &=& C_{1} \bigg(\frac{1 \mathrm{km}}{D}\bigg)^{p} \mathrm{for} D \leq 60 \mathrm{km},  \nonumber \\
N(>D) &=& C_{2} \bigg(\frac{1 \mathrm{km}}{D}\bigg)^{3} \mathrm{for} D > 60 \mathrm{km},
\label{e33}
\end{eqnarray}
where $C_{2}$ adopts a value of 7.9 $\times$ 10$^{9}$ and $C_{1}$ = $C_{2}$ (60)$^{p-3}$ by continuity for $D$ = 60 km.
The size distribution of Plutinos at small sizes is uncertain and one can find different proposed power law indexes in the 
literature. Then, following the analysis made by de El\'{\i}a et al. (\cite{deElia2008}) we use in our model 
three different initial populations, which are defined as follows
\begin{itemize}
\item Initial Population 1, with a cumulative power-law index $p$ of 3.0 for $D \leq$ 60 km,
\item Initial Population 2, with a cumulative power-law index $p$ of 2.7 for $D \leq$ 60 km,
\item Initial Population 3, with a cumulative power-law index $p$ of 2.4 for $D \leq$ 60 km.
\end{itemize}

\subsection{Collisional Parameters}

Here, we adopt constant values of the intrinsic collision probability $\langle Pi_{c} \rangle$
and the mean impact velocity $\langle V \rangle$ for Plutinos derived by Dell'Oro et al. (\cite{DellOro2001}).
Based on a sample of 46 Plutinos, these authors computed values of $\langle Pi_{c} \rangle$ and $\langle V \rangle$ 
of 4.44 $\pm$ 0.04 $\times$ 10$^{-22}$ km$^{-2}$ yr$^{-1}$ and 1.44 $\pm$ 0.71 km s$^{-1}$, respectively.

As for the impact strength, O'Brien \& Greenberg (\cite{OBrien2005}) showed that the general shape of the final
evolved asteroid population is determined primarily by the impact energy required for dispersal $Q_{D}$, but variations in
the shattering impact specific energy $Q_{S}$ and the inelasticity parameter $f_{ke}$ can affect such a final population
even if $Q_{D}$ is held the same. According to these arguments, we decide to choose a combination of the parameters
$Q_{S}$ and $f_{ke}$ that yield the $Q_{D}$ law from Benz \& Asphaug (\cite{Benz1999}) for icy bodies at 3 km\,\,s$^{-1}$.

de El\'\i a \& Brunini (\cite{deElia2007}) analyzed recently
the dependence of their numerical simulations on the shattering impact specific energy $Q_{S}$.
According to this work, the smallest gaps between
$Q_{S}$ and $Q_{D}$ curves lead to the smallest wave amplitudes in the size distribution of the final evolved
population as well as to the highest ejection rates of collisional fragments. Moreover, that study also indicates that
the formation of families is more effective for the simulations with a small gap between $Q_{S}$ and
$Q_{D}$ laws. Following these arguments, in this work we decide to use two $Q_{S}$ laws, $Q_{S,1}$ and $Q_{S,2}$, with a small and a large gap with
respect to the $Q_{D}$ law from Benz \& Asphaug (\cite{Benz1999}) for icy bodies at 3 km\,\,s$^{-1}$, respectively.
The results discussed in this work are those obtained using the $Q_{S,1}$ law. In Sect. 3.4, we develop numerical simulations using the 
$Q_{S,2}$ law in order to test the dependence of our results on such collisional parameter.
Figure~\ref{Fig1} shows the two $Q_{S}$ laws used in our simulations as well as the $Q_{D}$ law from
Benz \& Asphaug (\cite{Benz1999}) for icy bodies at 3 km\,\,s$^{-1}$.

Once the $Q_{S}$ law is specified, we fit the ineslaticity parameter $f_{ke}$
to get the Benz \& Asphaug (\cite{Benz1999}) $Q_{D}$ law.
According to O'Brien \& Greenberg (\cite{OBrien2005}), we express the parameter $f_{ke}$ as
\begin{equation} 
f_{\text{ke}} = f_{\text{ke}_{0}} \left(\frac{D}{1\,000\, \text{km}}\right)^{\gamma},
\label{e35} 
\end{equation}
where $f_{\text{ke}_{0}}$ is the value of $f_{\text{ke}}$ at 1\,000 km and $\gamma$ is a given exponent.
Our simulations indicate that the $Q_{D}$ law from Benz \& Asphaug (\cite{Benz1999})
for icy bodies at
3 km\,\,s$^{-1}$ is obtained with good accuracy from the combination of the selected
$Q_{S}$ law and $f_{\text{ke}}$, with $f_{\text{ke}_{0}} = 0.27$ and $\gamma = 0.7$. Such values are consistent with those from
Davis et al. (\cite{Davis1989}).

\subsection{Dynamical Considerations}

To study the orbital space occupied by the Plutino population,
we develop a numerical integration of 197 Plutino candidates extracted from the Minor Planet Center database with semimajor
axes between 39 and 40 AU. These objects are assumed to be massless particles subject to the gravitational field
of the Sun (including the masses of the terrestrial planets) and the perturbations of the four giant planets.
The simulation is performed
with the simplectic code EVORB from Fern\'andez et al. (\cite{Fernandez2002}). The
evolution of the test particles is followed for 10$^{7}$ years which is a timescale greater than any secular
period found in this resonance (Morbidelli \cite{Morbidelli1997}). From this, we build maps of the distribution of 
Plutinos in the orbital element planes ($a$,$e$) and ($a$,$i$), which allows us to determine the main stability regions
of the 3:2 Neptune resonance. Such maps are used to assign a characteristic orbit for every colliding Plutino and 
to specify the final fates of the different fragments generated in the collisional evolution.
A detailed description of this procedure can be found in de El\'\i a et al. (\cite{deElia2008}).

\section{Results}

The previously described collisional code allows us to calculate the collisional rates of Plutinos onto Pluto and Charon. From this,
it is possible to calculate the impactor flux of Plutinos of different sizes on the Pluto-Charon system. Besides, using a suitable expression
we can calculate the crater diameters produced by Plutinos on Pluto and Charon. We will present here our main results concerning the impactor 
flux and cratering onto Pluto and Charon due to the collisional evolution of the Plutino population.

\begin{figure}
\centering
\resizebox{\hsize}{!}{\includegraphics{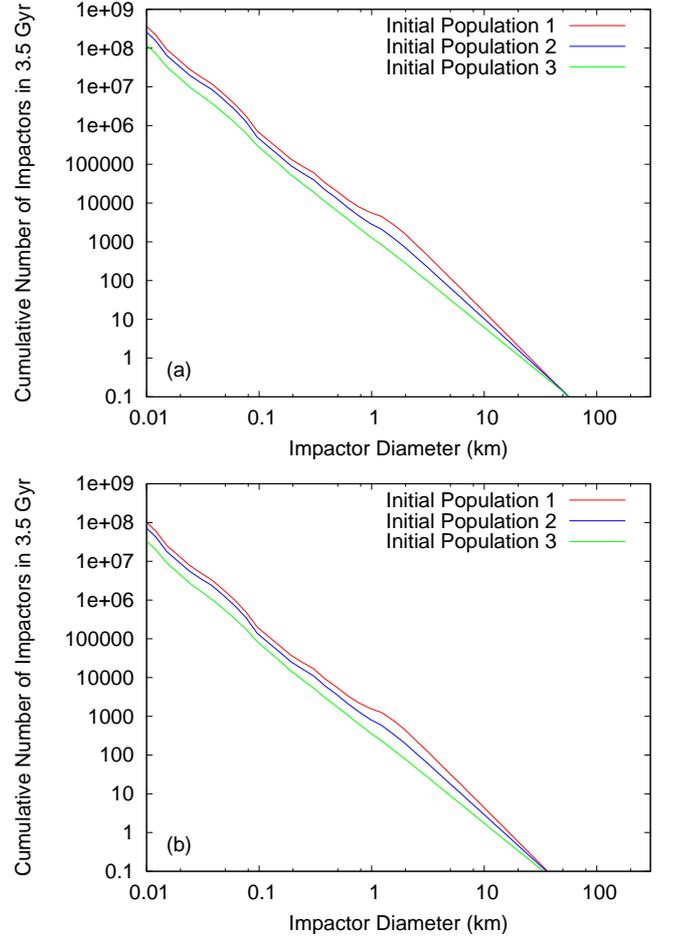}}
\caption{ Cumulative number of Plutino impactors onto Pluto a) and Charon b) over the last 3.5 Gyr as a function of impactor 
diameter.}
\label{fig2}
\end{figure} 

\begin{table}
\begin{minipage}[t]{\columnwidth}
\caption{
Collisional evolution results about Pluto.
 $N(D > 1)$  is the total number of  $D >$ 1 km Plutinos impacting Pluto over 3.5 Gyr,
 $\tau(D > 1)$  the time scale on which such impacts occur, $D_{\text{i,Max}}$ the 
largest Plutino expected to have struck Pluto during the past 3.5 Gyr, 
 $C(D > 1)$ the total number of  $D >$ 1 km craters on Pluto produced by 
Plutino impacts over 3.5 Gyr, and $D_{\text{c,Max}}$ the largest crater diameter on Pluto.
}
\label{Tabla1}
\centering
\renewcommand{\footnoterule}{}  % to avoid a line before footnotes
\begin{tabular}{l|c|c|c}
\hline \hline 
 & Initial Pop. 1   &  Initial Pop. 2   & Initial Pop. 3  \\
\hline 
 $N(D > 1)$  &   5\,552   &   2\,843   &   1\,271    \\
 $\tau(D > 1)$  (Myr)  &  0.63  &    1.23  &  2.75  \\
$D_{\text{i,Max}}$ (km) &  22.5  &    20.1  &   16.8   \\
$C(D > 1)$  &   113\,879   &   76\,726   &   43\,076    \\
$D_{\text{c,Max}}$ (km) &  38.8  &   35.5  &  30.1  \\
\hline
\end{tabular}
\end{minipage}
\end{table}

\begin{table}
\begin{minipage}[t]{\columnwidth}
\caption{
Collisional evolution results about Charon.
$N(D > 1)$  is the total number of  $D >$  1 km Plutinos impacting Charon over 3.5 Gyr,
$\tau(D > 1)$  the time scale on which such impacts occur, $D_{\text{i,Max}}$ the 
largest Plutino expected to have struck Charon during the past 3.5 Gyr, 
 $C(D > 1)$  the total number of $D >$  1 km craters on 
Charon produced by Plutino impacts over 3.5 Gyr, and
$D_{\text{c,Max}}$ the largest crater diameter on Charon.
}
\label{Tabla2}
\centering
\renewcommand{\footnoterule}{}  % to avoid a line before footnotes
\begin{tabular}{l|c|c|c}
\hline \hline 
 & Initial Pop. 1   &  Initial Pop. 2   & Initial Pop. 3  \\
\hline 
 $N(D > 1)$  &   1\,545   &   791    &  354    \\
 $\tau(D > 1)$  (Myr)  &  2.27  &  4.42  &   9.9 \\
$D_{\text{i,Max}}$ (km) &  14.6  &   12.5  &  9.8     \\
 $C(D > 1)$  &   50\,688   &   34\,541   &   20\,351    \\
$D_{\text{c,Max}}$ (km) &  32.7  &   28.9  &   24  \\
\hline
\end{tabular}
\end{minipage}
\end{table}

\subsection{Impactor Flux onto Pluto and Charon}

Figure~\ref{fig2} a) shows the  cumulative number of Plutino impacts onto Pluto over the last 3.5 Gyr as a function of impactor diameter, obtained
from the three different initial populations defined in Sect. 2.1.
Moreover, Table~\ref{Tabla1} summarizes some of our results concerning the impactor flux onto Pluto due to the collisional evolution 
of Plutinos.  From this, the number of  $D >$  1 km Plutinos striking Pluto over 3.5 Gyr is 
between 1\,271 and 5\,552, while the largest Plutino expected to have impacted Pluto during the past 3.5 Gyr had a diameter of 
 $\sim$  17 -- 23 km, depending on the initial size distribution.

On the other hand, 
Figure~\ref{fig2} b) shows the  cumulative number of Plutino impacts onto Charon over the last 3.5 Gyr as a function of impactor diameter.
Moreover, results about the impactor flux onto Charon due to the collisional evolution of Plutinos are summarized in Table~\ref{Tabla2}.
 From this, the number of $D >$  1 km Plutinos striking Charon over 3.5 Gyr is
between 354 and 1\,545, while the largest Plutino expected to have impacted Charon during the past 3.5 Gyr had a diameter of 
 $\sim$  10 -- 15 km, depending on the initial size distribution.

\subsection{Cratering on Pluto and Charon}

\begin{figure}
\centering
\resizebox{\hsize}{!}{\includegraphics{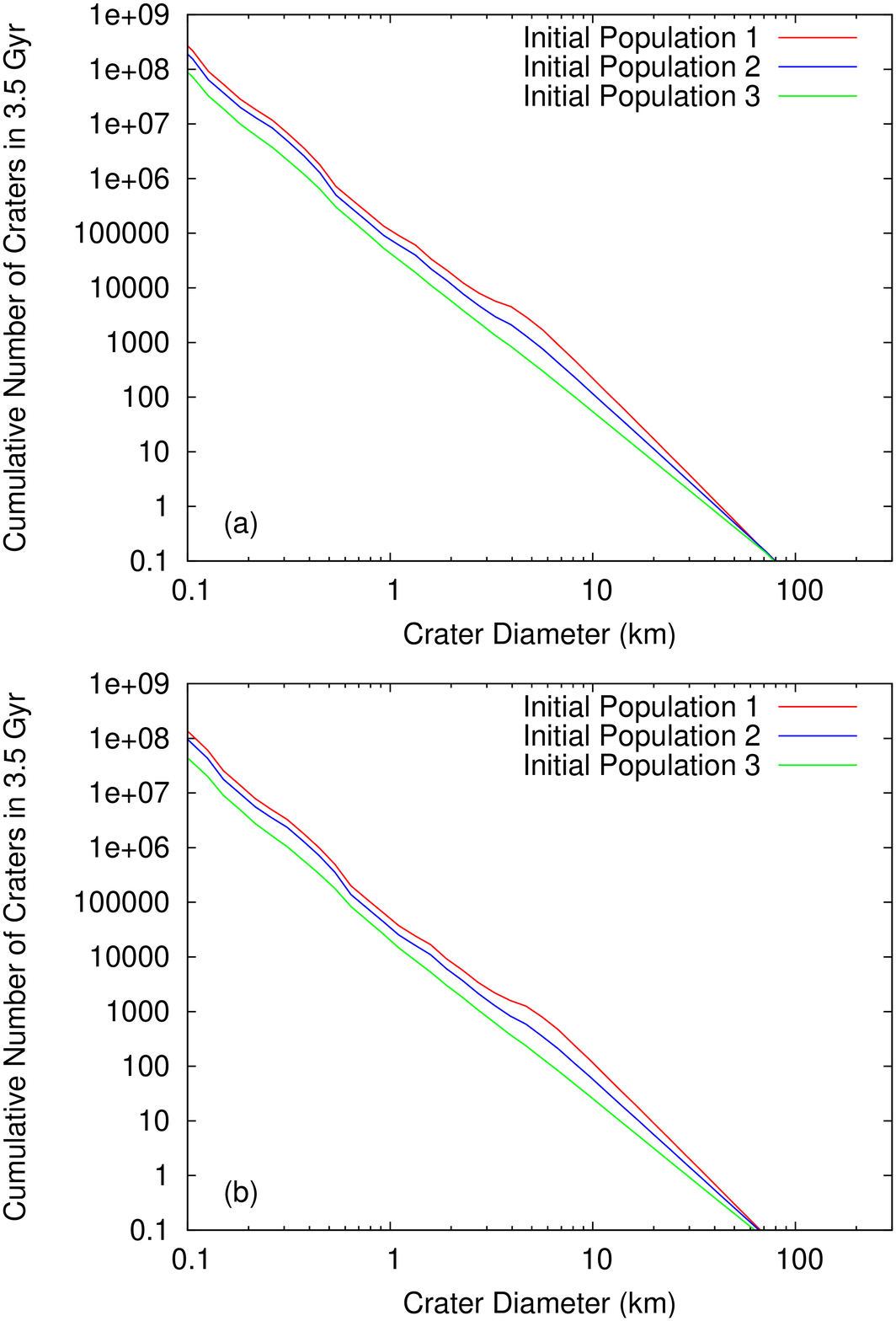}}
\caption Cumulative number of craters on Pluto a) and Charon b) produced by 
Plutinos over the last 3.5 Gyr as a function of crater diameter.
\label{fig3}
\end{figure}

Using the impactor fluxes previously estimated, it is possible to obtain the number of craters on Pluto and Charon over the last 3.5 Gyr as a function 
of crater diameter.
From Holsapple (\cite{Holsapple1993}), the diameter $D_{c}$ of a hemispherical crater is given by
\begin{equation}
D_{\text{c}} = 1.26 D_{\text{i}} (A \rho_{\text{i}}/\rho_{\text{t}})^{1/3} (1.61 g D_{\text{i}}/v_{\text{i}}^{2})^{-\alpha/3}
\label{ecuacion}
\end{equation}
where $D_{\text{i}}$ is the impactor diameter, $\rho_{\text{i}}$ the impactor density, $\rho_{\text{t}}$ the target density, 
$g$ the surface gravity of the target,
$v_{\text{i}}$ the encounter velocity, and $A$ and $\alpha$ are constants which depend on the mechanical properties of the target material. We adopt the
values of $A = 0.2$ and $\alpha = 0.6$ given by Holsapple (\cite{Holsapple1993}) for water 
ice.
The encounter velocity $v_{\text{i}}$ is given by
\begin{equation}
v_{\text{i}} = (U^{2} + v_{\text{esc}}^{2})^{1/2},
\end{equation}
where $U$ is the hyperbolic encounter velocity, which is equal to the mean impact velocity given in Sect. 2.2, and $v_{\text{esc}}$ is the escape 
velocity of the target. Values of the density, the surface gravity, and the encounter velocity used for Pluto and Charon are shown in 
Table~\ref{Tabla3}. 

Figure~\ref{fig3} a) shows the  cumulative number of craters on Pluto produced by 
Plutinos over 3.5 Gyr as a function of crater diameter.
 From this, the number of $D >$ 1 km craters on Pluto is 
between 43\,076 and 113\,879, while the largest crater diameter is between 
 $\sim$  31 and 39 km, depending on the initial size distribution.

At the same way, Figures~\ref{fig3} b) shows the  cumulative number of craters on Charon 
produced by Plutinos over 3.5 Gyr as a function of crater diameter.
 From this, the number of  $D >$  1 km craters on Charon is 
between 20\,351 and 50\,688, while the largest crater has a diameter of 
 $\sim$  24 -- 33 km, depending on the initial size distribution.

\begin{table}
\begin{minipage}[t]{\columnwidth}
\caption{
Values of the density $\rho$, the surface gravity $g$, and the encounter velocity $v_{i}$ used for Pluto and Charon.
}
\label{Tabla3}
\centering
\renewcommand{\footnoterule}{}  % to avoid a line before footnotes
\begin{tabular}{l|c|c}
\hline \hline 
 & Pluto   &  Charon  \\
\hline 
$\rho$ (g cm$^{-3}$) &  2.03   &  1.65     \\
$g$ (cm s$^{-2}$) & 66.4  &  27.9   \\
$v_{\text{i}}$ (km s$^{-1}$) & 1.9  &   1.6    \\
\hline
\end{tabular}
\end{minipage}
\end{table}

\subsection{Dependence of results on the number of Pluto-sized objects}

Kenyon et al. (\cite{Kenyon2008})
suggested that the resonant trans-Neptunian population has $\sim$ 0.01 -- 0.05 $M_{\oplus}$ in objects with $D \gtrsim$ 20 -- 40 km. From these 
estimates, it is possible to infer the existence of 5 Pluto-sized objects in the whole resonant population. If these 5 Pluto-sized objects
were all in the 3:2 Neptune resonance, we would have an upper limit for the large objects in this resonance.
Brown (\cite{Brown2008}), based on the completeness of the current surveys argued that two or three
more large KBOs are likely awaiting discovery. Since the existence of 5 Pluto-sized objects in the 3:2 Neptune resonance would seem to be
an overestimation, we assume an upper limit of 2 Pluto-sized objects in this resonant region in order to analyze the dependence of our 
simulations on the number of large objects in this resonance.

If 2 Pluto-sized objects are assumed in the 3:2 Neptune resonance, the total number of Plutino impactors both onto Pluto as Charon with 
diameters $D >$ 1 km is a factor of $\sim$ 1.6 -- 1.8 larger than that obtained considering only 1 Pluto-sized object in this resonant region. 

\subsection{Dependence of results on the shattering impact specific energy $Q_{S}$}        %$

To test the dependence of our results on the shattering impact specific energy $Q_{S}$, we carry out numerical simulations using the
$Q_{S,1}$ and $Q_{S,2}$ laws, 
which show a small and a large gap with respect to the $Q_{D}$ law from Benz \& Asphaug (\cite{Benz1999}) for icy bodies at 3 km\,\,s$^{-1}$,
respectively. We obtain that the $Q_{S,2}$ law produces a size distribution with a wave amplitude larger than that the one obtained using 
the $Q_{S,1}$ law as was noticed by de El\'{\i}a \& Brunini (\cite{deElia2007}). However, the whole impactor flux is almost preserved using
both laws.

\section{Discussion}

 Durda \& Stern (\cite{Durda2000}) developed a model of collision rates in the 
present-day Kuiper Belt and Centaur region. That is a static, multizone and 
multi-size-bin collision rate model that calculates instantaneous 
collision rates on Kuiper Belt objects, Centaurs and the Pluto-Charon 
system. 

The model of Durda \& Stern (\cite{Durda2000}) defines a disk in the 
30--50 AU zone whose surface mass density follows a profile of the form
$\Sigma \propto R^{-\beta}$, with 
$\beta =$  -2. Moreover, the disk adopts an average 
eccentricity  $\langle e \rangle$  and assumes an 
equilibrium condition where the average inclination  $\langle 
i \rangle = \frac{1}{2}\langle e \rangle$.  The values adopted for
 $\langle e \rangle$  are of 0.0256 and 0.2048,
which are representative of the values observed in the Kuiper Belt at that time.
Once the disk properties are specified, the disk is binned into a series of 
radially concentric tori 1 AU in width in each of which the size 
distribution of the population is represented by a two-component power 
law of the form  $N(D) \propto D^{b} dD$, where 
 $b = -3$  for  $D <$  10 km and 
 $b = -4.5$  for larger bodies.
  Durda \& Stern (\cite{Durda2000}) carried out their runs 
assuming the existence of 7 $\times$ 
10$^{4}$--1.4 $\times$ 10$^{5}$  objects with a radius $r >$ 
 50 km between 30 and 50 AU, according to the observational evidence from Jewitt et al. 
(\cite{Jewitt1998}) and Gladman et al. (\cite{Gladman1998}). 
%ver esta frase no  la entiendo a que viene
Moreover, they suggested that about 40 \% of the total population are 
in or near the 3:2 mean motion resonance with Neptune.
From this, the flux of  $r = 1$  km
 Kuiper Belt projectiles onto Pluto and Charon over 3.5 Gyr is found to be 
approximately 8.9 $\times$ 10$^{3}$  and 1.1 $\times$ 10$^{3}$, respectively.   

 However, the observational advances that have occurred during the last ten 
years have revealed a more complex dynamical structure of the trans-Neptunian 
region. In particular in the region between 30 and 50 AU there are very different dynamical  
classes of TNOs, like the classical, resonant, and scattered-disk populations. They  
show different dynamical features and ranges of orbital elements as 
well defined semimajor axis zones, 
eccentricities as high as  $\sim 0.45$  and inclinations as high as  $\sim 45^{\circ}$.
 Moreover, current data provide clear evidence for differences 
in the mass and size distribution parameters among such dynamical 
classes (Kenyon et al. \cite{Kenyon2008}). The results by Durda \& Stern (\cite{Durda2000})
are obtained for objects in that region, but they don't account for the different dynamical populations. 
Particularly, while the Plutino population is 
included quantitatively in Durda \& Stern's (\cite{Durda2000}) model, 
they did not account for the dynamical properties of such resonant 
population. Then, their results should be taken with caution.

\section{Conclusions}

We have presented a study aimed at analyzing the collisional
and dynamical evolution of the Plutinos. Assuming the
existence of one Pluto-sized object in the 3:2 Neptune resonance,
our main results are the following.

 Our results depend strongly on the initial size distribution of Plutinos. Depending on the initial population, we have obtained that the number of
 $D >$ 1 km Plutinos streaking Pluto over 3.5 Gyr is between 1\,271 and 5\,552. For Charon, the number of  $D >$ 
 1 km Plutino impactors is between 354 and 1\,545. Moreover, the 
number of  $D > 1$  km craters on Pluto produced by 
Plutinos during the last 3.5 Gyr is between 43\,076 and 113\,879. For 
Charon, the number of  $D > 1$  km craters is between 
20\,351 and 50\,688.
 On the other hand, the largest Plutino impactor onto Pluto has a diameter between 
 $\sim$  17 and 23 km, that produces a crater with a diameter of  $\sim$  31 -- 39 km. The largest Plutino impactor 
onto Charon has a diameter between  $\sim$ 10 and 15 km, that produces a crater with a diameter of  $\sim$  24 -- 33 km.

We have tested the dependence of results on the number of Pluto-sized objects in the Plutino population.  
The number of Plutino impactors on Pluto and Charon obtained considering 2 Pluto-sized objects is a factor
of $\sim$ 1.6 -- 1.8 larger than that obtained considering only 1 Pluto-sized object in this resonant region.

Besides, using two different $Q_{S}$ laws, with a small and a large gap with respect to the $Q_{D}$ law, we have obtained almost the same 
impactor flux onto Pluto and Charon. 

 The complex structure of the trans-Neptunian region, with its dynamically different populations, requires a  
detailed study of the contribution of all the potential sources of impactor on the Pluto-Charon system, to obtain
the main contributor and the whole production of craters. When Nasa's New Horizon Pluto-Kuiper Belt mission 
flies by Pluto system we will have images of the surface and cratering records 
could be measured.
Then, we will be able to contrast the theoretical studies of the production of craters with the
observations of the New Horizon mission which will help us to understand the geological processes and history of the surfaces 
of those worlds. 
In a future work we will  calculate the contribution of the different dynamical classes of the trans-Neptunian region to the 
flux of projectiles onto Pluto and Charon. From this, it would be possible to specify the 
primary source of impactors on such bodies and to determine if the 
Plutino population can be considered a significant source.

\end{document}